\documentclass[12pt]{iopart}
\usepackage[dvips]{graphicx}

\begin{document}

\title[Monte Carlo inspiraling rates]{Monte Carlo cluster simulations to
  determine the rate of compact star inspiraling to a central galactic
  black hole}
 
\author{Marc Freitag}
\address{Observatoire de Gen{\`e}ve, CH-1290 Sauverny, Switzerland}

\begin{abstract}
  The capture and gradual inspiral of stellar mass objects by a
  massive black hole at the centre of a galaxy has been proposed as
  one of the most promising source of gravitational radiation to be
  detected by LISA. Unfortunately rate estimates for this process
  suffer from many uncertainties. Here we report on the use of our
  newly developed Monte Carlo stellar dynamics code to tackle this
  problem. We present results from simple galactic nuclei models that
  demonstrate the high potential of our approach and point out the
  aspects of the problem where an improved treatment seems desirable.
\end{abstract}

\section{Introduction}

In the past decade, the harvest of observational evidences for the
presence of massive dark objects in the centre of most bright galaxies
has become very impressive. It is generally accepted that these mass
concentrations are most likely massive black holes (MBHs) with detected
masses ranging between $10^6$ and $10^{10}\,M_{\odot}$. 
For all but two galaxies, however, today's observations in the
electro-magnetic spectrum lack resolution to rule out concurrent
interpretations concerning the nature of the dark object
\cite{Ho99,Maoz98}.

The motion of a ``test particle'' (with mass $\ll M_{\mathrm{MBH}}$)
on a relativistic orbit, monitored by the emitted gravitational waves
(GW), is an ideal way to get much more precise information from the
immediate vicinity of the putative black hole. By detecting such
signals and comparing them with general relativity predictions, we
could, at the same time, test Einstein's theory in the strong field
regime, confirm the existence of MBHs (in a statistical sense, if not
in particular galaxies) and get precise determinations of their mass
and spin \cite{Thorne98}.

However, two main classes of theoretical difficulties have to be worked
around. First, the computation of the trajectory of a particle around 
a Kerr BH without resorting to strong simplifying assumptions, is a
formidable, still unsolved, problem.
The second difficulty is the prediction of the number of potential
sources, i.e. the rate with which stars get so deeply bound to a
central MBH that their subsequent orbital evolution is driven by
GW-emission.

The range of MBH masses that can be probed by LISA is $3\times
10^5-3\times 10^7\,M_{\odot}$ \cite{Thorne98}.  Only for the highest
$M_{\mathrm{MBH}}$ values can MS stars be captured on relativistic
orbits without being teared off by tidal forces so we expect most
capture events to feature a compact remnant.
It can be brought onto relativistic orbits through 2-body relaxation
or non disruptive collisions with other stars if the velocity
dispersion exceeds $V_{\ast}$, an unlikely situation for compact
stars.
Repeated impacts on an accretion disk could also channel stars onto
relativistic orbits \cite{SyerClarkeRees91}, but this process is
probably ineffective for compact remnants as they would not sweep much
disk's material at each crossing.
 
The rate of captures on orbits with strong emission of gravitational
radiation (hereafter ``GR-captures'') has been estimated
in three previous works
\cite{HilsBender95,SigurdssonRees97,MiraldaGould00}.
Unfortunately, these authors use
different galactic nuclei models, assumptions and computation
methods\ldots\ and quite expectedly get discordant capture rates,
ranging from $2\times 10^{-8}$ to $10^{-4}$\,yr$^{-1}$.
Besides the determination of such a rate, important issues
raised by these works are the following. To what extent is the
mass-segregation process efficient at concentrating stellar remnants
towards the central regions? Is the GR-capture rate sustainable for
many Gyrs? How does the stellar nucleus evolve on the long run in
response to this and other stellar dynamical processes (relaxation,
tidal disruptions, collisions, stellar evolution)?
 
\section{The Monte Carlo cluster evolution code}
\label{sec:MC_code}

\subsection{Overview of the code}
\label{subsec:MC_overview}
We recently developed a new computer code to follow the evolution of a
central star cluster surrounding a MBH over time scales as long as
$10^{10}$\,years. This tool, which we describe in
\cite{FreitagBenz00a}, is the backbone of our work aimed at a better
modelization and understanding of the influence of the MBH on the
stellar system.

The numerical scheme we chose is a Monte Carlo (MC) method based on
the pioneering work of H{\'e}non \cite{Henon73}. This class of
programs proved very successful in the realm of globular clusters
studies \cite{Stodolkiewicz82,Giersz98,Joshi00}.  In its basic form,
the method is more or less equivalent to resolving the Fokker-Planck
(FP) equation, but with no implicit integration of the diffusion
coefficients. The relaxation-driven secular evolution of a stellar
system is simulated without wasting CPU time at integrating orbits.
Compared to direct FP resolutions, the main advantages of the MC
method are {\bf (1)} its simplicity, {\bf (2)} the fact that any
stellar mass function and any velocity distribution (both variable in
time and position) are naturally coped with, as is the cluster's
self-gravity, and {\bf (3)} the ease with which extra physical
processes can be included, even those that are not continuous, like
collisions.  Although they are also particle-based, MC simulations are
enormously faster than relaxational $N$-body integrations whose
computing time increases like $N_{\mathrm{part}}^{2-3}$ while the
scaling is almost linear for our MC code.
Furthermore, as each particle represents a spherical shell of stars
with same orbital and stellar properties, the number of simulated
stars can be made arbitrarily high.
Of course, MC methods also suffer from a few drawbacks. On the one
hand, they produce much noisier results than FP methods. On the other
hand, they rely on more simplifying assumptions than $N$-body schemes;
the strongest of them are spherical symmetry, constant dynamical
equilibrium and treatment of relaxation as the
integrated effect of distant, uncorrelated, 2-body encounters.
 
As far as we know, our code is the first implementation of the
H{\'e}non's scheme devoted to MBH-hosting dense galactic nuclei. We
spent much time to include stellar collisions (between MS stars,
for the time being) as realistically as possible
\cite{FreitagBenz00b,FreitagBenz00c}.  From more than $12\,000$ MS-MS
collision simulations carried out with a Smoothed Particle
Hydrodynamics code (SPH, \cite{Benz90}), we built a grid of
collisional outcomes from which the result of any collision is
interpolated when it occurs during a cluster evolution simulation. We
also included ``loss-cone'' effects \cite{LightmanShapiro77} to
account for the tidal disruptions of stars in the Roche zone around
the MBH. Finally, for this LISA symposium, we adapted this loss-cone
routine in order to detect GR-capture events.
 
\subsection{Computing capture rates with the Monte Carlo code}
\label{subsec:MC_rates}

Following Sigurdsson \& Rees \cite{SigurdssonRees97}, we consider that
a star is ``captured'' by the MBH if it gets on an orbit whose
shrinkage time by GW-emission, $T_{\mathrm{GW}}$, is shorter than the
time 2-body relaxation would take to modify it substantially,
$T_{\mathrm{mod}}$. The most likely way for a star to get sufficiently
close to the MBH to be captured is to experience 2-body encounters
that make its orbit nearly radial. For such a highly eccentric orbit,
$T_{\mathrm{mod}}$ is the time taken by relaxation to introduce a
relative change in the pericentre distance of order 1. 
Then, $T_{\mathrm{mod}} \simeq (1-e)T_{\mathrm{relax}} \ll
T_{\mathrm{relax}}$ where $e$ is the eccentricity. Typical values of
$1-e$ for captured particles are $10^{-7}-10^{-3}$.

Using time steps $\delta t$ as small as a fraction of
$T_{\mathrm{mod}}$ would completely jeopardize the efficiency of the
MC scheme for which $\delta t \simeq 0.001-0.01\,T_{\mathrm{relax}}$ is
otherwise sufficiently small. For a star with fixed $|\vec{V}|$ at a
given distance from the centre, the capture orbits form a thin conical
bundle with aperture angle $\theta_{\mathrm{GRC}} \ll \pi$. If we
impose too large a $\delta t$, $\vec{V}$ jumps over this tiny ``loss
cone'' and most captures are missed. This problem is closely
reminiscent to the detection of tidal disruptions. Thus we solved it
the same way.  We keep time steps that are a (small) fraction of
$T_{\mathrm{relax}}$ (and/or $T_{\mathrm{coll}}$) but we
``over-sample'' each of them by simulating the random walk of the tip
of $\vec{V}$ on a sphere during $\delta t$.  We simply piggy-backed
the procedure for GR-captures detections on the routine for tidal
disruptions. Hence, as a star would be tidally destroyed only if it
stays on a loss cone orbit until next pericentre passage, the time
resolution of the random walk is the orbital period
$T_{\mathrm{orb}}$. This is not well adapted to GR-captures for which
$T_{\mathrm{orb}}$ has not special meaning.  However, entries in the
GR-capture loss cone are correctly simulated as a diffusion process as
long as individual steps in the random walk are smaller than
$\theta_{\mathrm{GRC}}$. But this is always true because, by
definition, the R.M.S. diffusion angle during $T_{\mathrm{GW}}$ is
$\theta_{\mathrm{GRC}}$ and $T_{\mathrm{GW}}$ is much longer than
$T_{\mathrm{orb}}$.

We think that the main intrinsic limitation in our approach is to
treat relaxation with small angle approximation. Indeed, close
gravitational encounters with scattering angles of order $\pi$,
although they only contribute a fraction $\ln(\Lambda)^{-1}<0.1$ to
the overall relaxation, 
can
dominate the rate of captures \cite{SigurdssonRees97}.  Another
limitation is of statistical nature.  To ensure energy conservation,
each particle stands for a fixed number of stars. Consequently there
are only 1300 SBH particles out of $2\times 10^6$.

\section{Simulations of simple models of galactic nuclei}

\subsection{Initial model \& included physics}
To explore the potential of our MC code in predicting GR-capture
rates, we run a few simulations that are variations around our
``standard'' galactic nucleus model. The stellar cluster is set as a
$W_0=8$ King model, with a core radius of 0.47\,pc and $3.6\times
10^8$ stars.
As stellar evolution is not simulated, we have to include an evolved
stellar population, containing compact remnants, from the beginning of
the computations.  It is prepared according to the prescription of
Miralda-Escud{\'e} \& Gould \cite{MiraldaGould00}. The number fractions
of white dwarfs (WDs), neutron stars (NSs) and SBHs are 0.056,
$3.5\times 10^{-3}$ and $6.5\times 10^{-4}$, and their
individual masses 0.6, 1.4 and 7.0\,$M_{\odot}$, respectively. We do not
include giant stars. An initial ``seed'' black hole
($M_{\mathrm{BH}}/M_{\mathrm{clust}}\simeq 10^{-4}$) is set at the
centre. It accretes all gas released by stellar collisions and tidal
disruptions with no delay. GR-captured stars are also assumed to be
immediately swallowed by the BH.  Direct plunges through the
horizon are detected. Collisions between MS stars are treated
realistically thank to our SPH grid but tidal disruptions are assumed
to be always total.

This model is not meant to be highly realistic. It is probably too
dense and massive to represent a real galactic nucleus. To bracket the
density value at the Galactic centre \cite{Genzel96}, we also used
another model with a number of stars reduced by a factor 10.  A better
representation for a galactic nucleus would require a power law
density profile at large radii ($\rho\propto R^{-\alpha}$ with
$\alpha\leq 2$ typically) instead of a steep cut-off.  This point
is problematical as it imposes to put large amounts of particles at
large distances and consequently reduce the resolution near the MBH
where all the action takes place.

\subsection{Simulation results}

\begin{figure}
  \resizebox{\hsize}{!}{\includegraphics{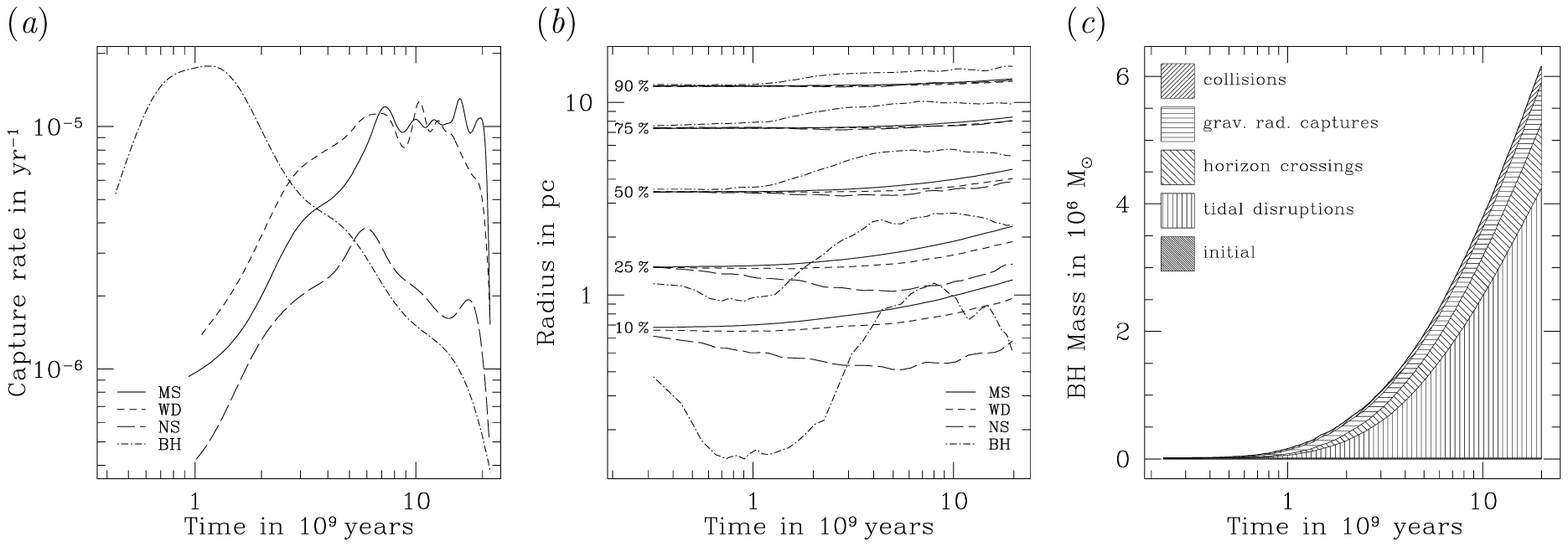}}
  \caption{
    Evolution of the ``standard'' model (see text). Panel ({\it a})
    shows the rate of captures by emission of gravitational radiation
    for various stellar types.  Panel ({\it b}) depicts the evolution
    of a few Lagrangian radii.  Panel ({\it c}) shows the contribution
    of various processes to the growth of the central BH.}
  \label{fig:evol1}
\end{figure}

\begin{figure}
  \resizebox{\hsize}{!}{\includegraphics{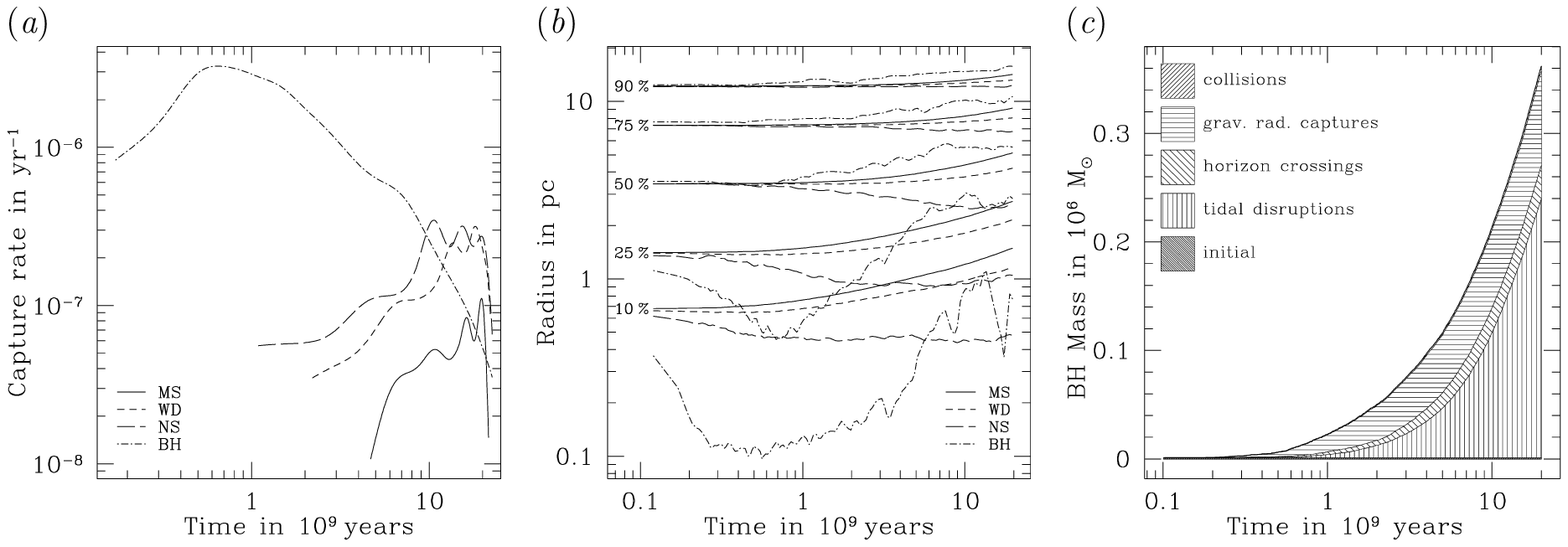}}
  \caption{
    Same as figure~\protect\ref{fig:evol1} for a nucleus model which is
    10 times less massive. }
  \label{fig:evol2}
\end{figure}

Some aspects of the evolution of the ``standard'' model are shown in
figure~\ref{fig:evol1} while figure~\ref{fig:evol2} illustrates the
evolution of the lighter nucleus. Both simulations were realized
with $2\times 10^6$ particles.

We see that relaxational segregation of SBHs towards the centre occurs
quickly. An early broad peak in the capture rate for this species
ensues.  $\dot{N}_{\mathrm{SBH}}$ then steadily decreases due to
exhaustion of SBHs in the central regions.  At late times, the capture
rate is dominated by WDs, NSs, and, for dense clusters, low-mass MS
stars ($\langle M_{\ast} \rangle \simeq 0.2\,M_{\odot}$). However, we
noted that for many of these MS stars, we have $T_{\mathrm{GW}}\gg
10^9\,$years even though the capture condition
($T_{\mathrm{GW}}<T_{\mathrm{mod}}$) is obeyed. Further investigations
about these cases are called for. Typical remnant capture rates at
$T=1-2\times 10^{10}$\,years are $\sim
10^{-5}\,\mathrm{yr}^{-1}$ and $\sim5\times 10^{-7}\,\mathrm{yr}^{-1}$
for the standard and ``light'' models. A nucleus with 10 times the
mass of the standard model yields $\sim 10^{-4}\,\mathrm{yr}^{-1}$.

A massive BH ($M>10^{6}\,M_{\odot}$) can be grown from a seed if the
initial stellar density is high enough. However, the stellar density
after a Hubble time remains uncomfortably high ($\sim
10^7\,M_{\odot}\mathrm{pc}^{-3}$ at 0.1\,pc). Tidal disruptions are
the major contributors to the BH's mass with stellar collisions
playing only a minor role.
In the lighter
model, GR-captures of SBHs dominates the early growth of the central
BH with tidal disruptions catching up later on. This is probably
connected to the stronger relaxation in this model.

\subsection{Future improvements}
There are many ways our simulations could be improved in the
future. Let's mention a few of them that are relevant for the problem
of GR-captures.
\begin{itemize}
\item The lack of resolution for rare species, such as SBHs, and the
  problem of extending the initial model to larger radii could be
  solved if each particle could represent a different number of stars.
  We presently do not know how to achieve this. Alternatively, a
  massive increase in the number of particles would solve these
  difficulties. A simulation with $10^7$ particles would require $\sim
  1$~month of CPU time, but we run short of computer memory. To go
  beyond this number, a parallel version of the code should be
  developed.
\item Stellar evolution should be included. In taht case, we have to
  determine the fate of the gas lost by stars, and particularly the
  fraction accreted onto the central BH. For our IMF, more than 40\,\%
  of the ZAMS mass is lost from the stars in the first
  $10^{10}$~years, which amounts to $5.9 \times 10^7\,M_{\odot}$ for
  the standard model!  Preliminary calculations including stellar
  evolution indicate that the GR-capture rate at late stages is not
  substantially altered when all the emitted gas is lost from the
  nucleus.
\item A better treatment of GR-capture events should be attempted. It
  could go beyond the simple $T_{\mathrm{GW}}<T_{\mathrm{mod}}$
  criterion and the assumption of immediate swallowing. 
\item The contribution of large angle scatterings to GR-captures must
  be clarified and included in our simulations if important.
\end{itemize}

\ack

The author wants to heartily thank Dr.~Sterl Phinney for the invitation 
to the 3$^\mathrm{rd}$ LISA Symposium which gave him the opportunity
to peek at the fascinating world of gravitational waves.

\end{document}